\documentclass[pre,twocolumn,showpacs,floatfix]{revtex4}
\usepackage{graphicx}
\begin{document}

\title{Role of reversibility in viral capsid growth: A paradigm for
self-assembly}
\author{D. C. Rapaport}
\email{rapaport@mail.biu.ac.il}
\affiliation{Physics Department, Bar-Ilan University, Ramat-Gan 52900, Israel}
\date{September 7, 2008}

\begin{abstract}
Self-assembly at submicroscopic scales is an important but little understood
phenomenon. A prominent example is virus capsid growth, whose underlying
behavior can be modeled using simple particles that assemble into polyhedral
shells. Molecular dynamics simulation of shell formation in the presence of an
atomistic solvent provides new insight into the self-assembly mechanism, notably
that growth proceeds via a cascade of strongly reversible steps and, despite the
large variety of possible intermediates, only a small fraction of highly bonded
forms appear on the pathway.
\end{abstract}

\maketitle

Structure formation in self-assembling molecular complexes is a particularly
important process, both for nature and nanotechnology \cite{whi02}. One very
familiar example is the formation of viral capsids, the polyhedral shells of
capsomer particles enclosing the genetic payload of spherical viruses
\cite{cri56,cas62}. The ubiquity of icosahedral symmetry among spherical
viruses, a design adopted by nature where shells are built by repeated use of
just one or a small number of distinct capsomers, motivates a reduced
description that avoids the molecular details of capsomer proteins; the fact
that assembly also occurs {\em in vitro} \cite{pre93,zlo99,cas04} makes it an
ideal candidate for simulation. If self-assembly is indeed governed by general
organizational principles -- with important consequences for both medicine and
materials science -- they ought to be accessible using simplified models.

There is little direct experimental evidence on the nature of assembly pathways,
and since self-assembly implies a nonequilibrium state, where predictive theory
is absent, simulation has a potentially important role. Molecular dynamics (MD)
modeling of capsid self-assembly, based on simple structural models, was
described in Refs.~\cite{rap99,rap04}; the model particles retain just enough
detail to ensure meaningful behavior, the two key features being an effective
molecular shape formed out of soft spheres rigidly packed so particles fit
together in a closed shell, and multiple interaction sites positioned to
stabilize the correct final structure. The focus was on achieving assembly;
pathways were not considered, and solvent was omitted to reduce computational
cost.

The present paper describes MD simulations of self-assembling particles that
incorporate an explicit atomistic solvent. Solvent presence aids cluster breakup
without subassemblies needing to collide directly, curtails the ballistic nature
of the particle motion ensuring conditions closer to equilibrium, serves as a
heat bath to absorb energy released by bond formation, and maintains the
dynamical correlations of a fluid medium. To anticipate the conclusions,
self-assembly is found to consist of a cascade of reversible stages, with a
strong preference for low-energy intermediate states, eventually leading to a
high yield of complete shells. Paradoxical though it may seem, reversibility is
the key to efficient production. Even though the present focus is on icosahedral
growth -- to lower the computational effort -- these features of the behavior
are likely to be entirely general.

An alternative particle-based, solvent-free simulation \cite{ngu07} treated
quasi-rigid bodies made of hard spheres. Even simpler capsomer representations
have been proposed with spherical particles instead of extended capsid shapes,
and either directional interactions \cite{hag06} whose range exceeds the
particle size, or bonding energies determined by local neighborhood rules
\cite{sch98}; solvent is represented implicitly with stochastic forces. At the
other extreme are the folded proteins of real capsomers; MD treatment of
all-atom models \cite{fre06} is limited to short trajectories to test stability
of prebuilt shells. Capsid structure has also been studied by a variety of
theoretical methods \cite{lid03,twa04,zan04,hic06,hem06}, and experiments have
been interpreted using concentration kinetics \cite{zlo99,van07}; all avoid
addressing the underlying discrete particle dynamics.

\begin{figure}
\includegraphics[scale=0.45]{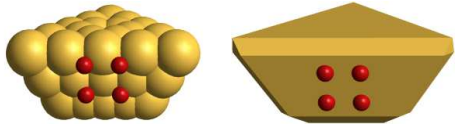}
\caption{\label{fig:1} (Color online) Model particle and the effective truncated
pyramidal shape; small spheres denote interaction sites.}
\end{figure}

The simulations involve triangular particles based on the design and
interactions of Ref.~\cite{rap04}, that are expected to self-assemble into
icosahedral shells. Particle shape is approximated by a rigid set of soft
spheres with lateral faces inclined at $20.905^\circ$ to the normal, as shown in
Fig.~\ref{fig:1}. There are four interaction sites on each lateral face that can
bond to matching sites on adjacent particles; multiple sites help maintain
correct alignment after bonding. The solvent atoms are identical to the particle
spheres; all experience a soft-sphere repulsion whose parameters determine the
dimensionless MD length and time units \cite{rap04bk} used here. The attractive
interaction between bond-forming interaction sites is harmonic at distances
below $r_h=0.3$ and inverse-square above $r_h$ with range $r_a=3$; the overall
attraction strength is governed by a parameter $e$. The fact that particle size
exceeds interaction range, although less so than in real capsomers, reduces the
interaction of wrongly positioned particles. 

MD methodology is described in Ref.~\cite{rap04bk}. The system consists of
125\,000 molecules in a cubic region with periodic boundaries; most are solvent
atoms, but there are 1875 triangular particles (1.5\%), enough to produce 93
full shells; the mean number density is 0.2. Although particle concentration is
much higher than in experiment, the solvent presence is sufficient to ensure
that diffusion tempers the otherwise ballistic particle motion. Runs of 60
million time steps are typically needed, at a rate of one million steps per day
on a computer with dual 3.6GHz Intel processors; there are 200 steps per unit
(MD) time. Particle mass, proportional to volume, is 21 times the solvent atom
(with unit mass); having a much smaller ratio than in actual viruses leads to a
reduced assembly timescale, accessible to MD, without qualitative change in
behavior. Gradual heating by exothermal bond formation is suppressed by means of
a thermostat that maintains a temperature of 0.667, corresponding to unit mean
kinetic energy. Runs are initialized by placing particles and solvent atoms on a
lattice with random velocities; to avoid overlap, particles begin collapsed and
expand to their final shape over the initial 5000 steps. No additional
mechanisms are included, e.g., damping and partial shell breakup \cite{rap04},
to aid or regulate assembly.

The ability to produce complete shells is the principal characteristic of the
method; since $e$ is the only parameter varied, coverage of the phase diagram is
limited, but this proves adequate for demonstrating a variety of growth
scenarios. Establishing shell completeness requires identifying bound clusters
\cite{rap04bk} and checking the bond network connectivity. There is a certain
arbitrariness in the bond definition. Interaction sites are bonded when less
than 0.6 ($= 2 r_h$) apart, a threshold that avoids transient bond breakage by
thermal vibration. Particles are considered bonded if all four mutual site pairs
are bonded, a state implying alignment; a weaker condition involving just a
single site pair is used to search for loosely linked, misaligned clusters.

\begin{figure*}
\includegraphics[scale=0.90]{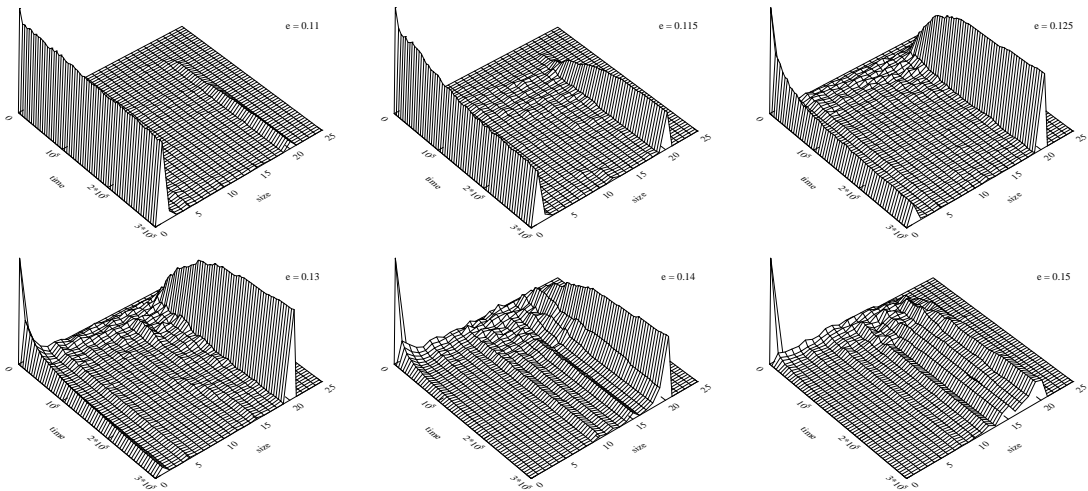}
\caption{\label{fig:2} Cluster size distributions as functions of time (MD
units); the distributions, including monomers, are expressed as mass fractions,
and $e$ (attraction strength) values are selected to show the different growth
scenarios.}
\end{figure*}

Figure~\ref{fig:2} shows how the cluster size distributions, expressed in terms
of mass fractions, evolve with time for particular $e$ values. The highest
observed shell count, 83, occurs at $e=0.13$ and corresponds to a yield of 89\%.
Growth curves have a sigmoidal shape; there is an initial lag until shells
appear, followed by a period of rapidly increasing shell count that ends
asymptotically. Further change, beyond the 60 million steps shown, is extremely
slow. No oversized clusters are observed (although sufficiently large $e$ would
lead to mutant forms).

\begin{figure}
\includegraphics[scale=0.65]{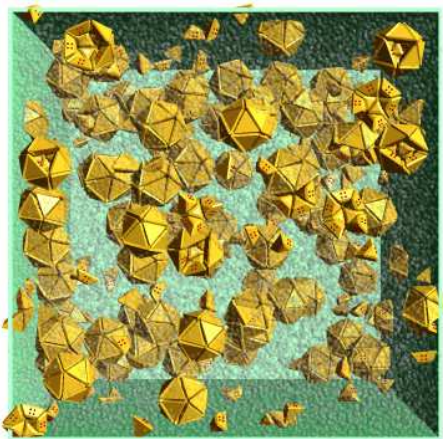}
\caption{\label{fig:3} (Color online) Snapshot of the $e=0.13$ system with 80
complete shells; solvent atoms are shown semi-transparently, and there are
visual artifacts (closed shells that appear open and particles protruding
through walls) due to periodic boundaries.}
\end{figure}

There is a range of $e$ (exemplified by $e=0.13$) with efficient shell
production, where small clusters grow to completion while maintaining an
adequate monomer supply; the details depend on $e$. At lower $e$ (0.11) there
is practically no growth due to minimal initiation. At higher $e$ (0.14, and
especially 0.15) there is excessive early growth, resulting in too many monomers
being incorporated into clusters prematurely, and preventing the appearance of
complete shells until monomers are released by cluster breakup. The monomer
disappearance rate increases with $e$, unlike the more complex $e$-dependent
shell growth; for smaller $e$, a finite monomer fraction persists. While similar
overall behavior is seen in reaction kinetics studies \cite{zlo99}, provided
nucleation is rate limited, MD needs no such restriction. Reproducibility is
confirmed by repeating a run with a different initial state and obtaining
similar results. Figure~\ref{fig:3} provides a view of the $e=0.13$ system once
80 complete shells have formed.

Any tendency to form loosely linked clusters can be detected by relaxing the
bonding criterion to require only a single interaction site pair. Cluster
distributions are barely affected since bonds that persist eventually become
part of a strong (four-pair) bond. Direct visualization reveals that adjacent
partial shells can appear to be a single oversized cluster, but their weak
binding allows them to separate with little impact on long-term growth. Shell
stability is tested by extending a run that had already produced numerous closed
shells, after reducing $e$ to a value where assembly yields only a few dimers.
Residual small clusters promptly vanish, followed by the gradual disappearance
of larger clusters; eventually just the original closed shells remain, together
with the occasional dimer. This implies hysteresis \cite{sin03}, the survival of
complete shells even when conditions become unfavorable.

\begin{figure}
\includegraphics[scale=0.60]{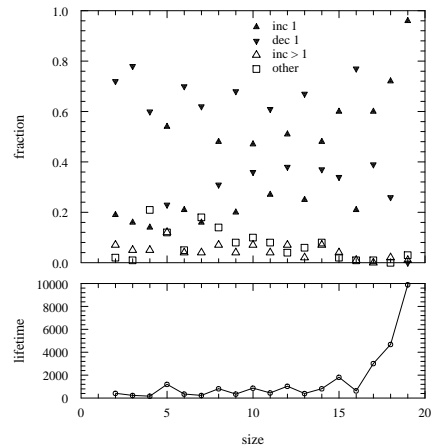}
\caption{\label{fig:4} Fraction of events producing unit size increases and
decreases, increases greater than unity, and others; mean cluster lifetimes (MD
units).}
\end{figure}

Other aspects of cluster development require further analysis. While exhaustive
shell histories are informative but not readily summarized, the statistics of
event types as cluster membership changes provide insight into the process.
Figure~\ref{fig:4} shows event fractions and mean cluster lifetimes (lifetimes
have broad distributions) for all sizes. The analysis considers particle
configurations during the first 30 million steps of the run with the highest
shell yield ($e=0.13$), spaced every 2000 steps to minimize merged (e.g., two
added monomers that appear as dimer addition) and missed (e.g., bonds that form
and break) events; clusters present at the end are excluded. The principal event
types are increases and decreases of unit size (including dimer breakup), and
size increases $>1$; others less readily specified (e.g., size decreases $>1$)
are grouped together (early in the run, there is slightly more dimer and trimer
growth, and direct visualization reveals infrequent events such as pentamer
bonding).

Reversibility plays a central role; for sizes $<19$ there is considerable
dissociation, and for half the sizes breakup is even more likely than growth.
The fact that growth events are only a fraction of the total enhances the
assembly process by providing ample opportunity for the error correction needed
to avoid kinetic traps. Reversibility also reflects a proximity to equilibrium,
with a slight bias in favor of growth. The consequences of reversibility have
also been considered \cite{zlo94} using reaction kinetics.

Insofar as lifetimes are concerned, pentamers live much longer than other small
clusters; lifetime and growth probability appear correlated. Nearly complete
shells live longest, but even at size 19 the wait until subsequent shell closure
is only eight times the pentamer lifetime. Certain growth steps allow incoming
particles to bond just along one face, a state favoring rapid disassembly (as
in, e.g., hexamers), but even more stable intermediates can lose members. Other
results (not shown) are that while growth probability and lifetime vary with
$e$, the overall trends are similar, and that mean bond length ranges from 0.2
for dimers to 0.05 for closed shells.

\begingroup
\squeezetable
\begin{table}
\caption{\label{tab:1} Intermediate states along the growth pathways: mean
cluster fractions ($f$) grouped by size ($s$) and bonds ($b$); numbers of
distinct cluster realizations ($n$) are included for comparison, with the final
columns enumerating other realizations that are not observed (sizes with unique
bond counts are omitted).}
\begin{ruledtabular}
\begin{tabular}{r@{\ \ \ \ \ }rrcrrcrrc@{\ \ \ \ \ \ \ \ }rr}
      & \multicolumn{9}{c}{Observed} &         \multicolumn{2}{c}{Others}  \\
  $s$ &$b\ $ &$n$& $f$  & $b\ $ &$n$ & $f$ &$b\ $& $n$ & $f$   & $b\ $  & $n$ \\
\hline
   5  &  5: & 1 & 0.948 &  4: &  5 & 0.052 &     &     &       &        &     \\
   6  &  6: & 1 & 0.953 &  5: & 13 & 0.047 &     &     &       &        &     \\
   7  &  7: & 4 & 0.979 &  6: & 22 & 0.021 &     &     &       &        &     \\
   8  &  9: & 1 & 0.851 &  8: & 11 & 0.140 &  7: &  46 & 0.009 &        &     \\
   9  & 10: & 3 & 0.938 &  9: & 27 & 0.062 &     &     &       &     8: &  79 \\
  10  & 12: & 1 & 0.808 & 11: & 13 & 0.166 & 10: &  60 & 0.026 &     9: & 151 \\
  11  & 13: & 3 & 0.931 & 12: & 28 & 0.069 &     &     &       & 11-10: & 328 \\
  12  & 15: & 1 & 0.917 & 14: & 11 & 0.073 & 13: &  74 & 0.010 & 12-11: & 446 \\
  13  & 16: & 4 & 0.876 & 15: & 31 & 0.105 & 14: & 142 & 0.019 & 13-12: & 372 \\
  14  & 18: & 1 & 0.802 & 17: & 15 & 0.198 &     &     &       & 16-13: & 417 \\
  15  & 20: & 1 & 0.825 & 19: &  5 & 0.146 & 18: &  38 & 0.029 & 17-15: & 170 \\
  16  & 21: & 4 & 0.915 & 20: & 19 & 0.068 & 19: &  38 & 0.017 &    18: &  28 \\
  17  & 23: & 1 & 0.923 & 22: &  7 & 0.077 &     &     &       &    21: &  12 \\
  18  & 25: & 1 & 0.888 & 24: &  5 & 0.112 &     &     &       &        &     \\
\end{tabular}
\end{ruledtabular}
\end{table}
\endgroup

The nature of the intermediate states along the growth pathways is an especially
notable feature. Table~\ref{tab:1} summarizes a series of measurements at half
million step intervals, for $e=0.13$, in which clusters are grouped by bond
count (other aspects of cluster geometry are not considered). There is a strong
preference for maximally bonded (minimal energy) clusters, and all are within
two bonds of maximum. Numbers of possible cluster realizations -- equivalent to
the distinct connected embeddings \cite{rap87} of triangles in an icosahedron --
are included for contrast. The majority are loosely bound and never observed as
intermediate states; e.g., while 91.7\% of clusters of size 12 are seen to adopt
the unique 15 bond form, and the remainder have 14 or 13 bonds, none of the 446
possible realizations with fewer bonds are encountered. The effect of an imposed
preference for maximally bonded intermediates has been studied with reaction
kinetics \cite{end05}; in MD this property emerges naturally from the
simulations.

In conclusion,
self-assembly at submicroscopic scales, where atomistic effects become
important, is very different from inherently unidirectional macroscopic
assembly. MD simulation reveals reversibility along the assembly pathway, with
dissociation often more likely than growth. Reversibility diminishes the
significance of kinetic traps because escape is accomplished by dissociation.
The coexistence of reversibility and a high error-free yield is a result likely
to have important implications for understanding supramolecular assembly in
general and capsid formation in particular.

The author is grateful to A. Zlotnick for helpful discussion.

\bibliography{selfassem}

\end{document}